\documentclass[aps,pre,twocolumn,groupedaddress,showpacs]{revtex4}
\usepackage{graphicx}
\usepackage{amssymb}

\begin{document}

\title{The partition of energy for air-fluidized grains}


\author{A.R. Abate and D.J. Durian}
\affiliation{Department of Physics \& Astronomy, University of
Pennsylvania, Philadelphia, PA 19104-6396, USA}


\date{\today}

\begin{abstract}

The dynamics of one and two identical spheres rolling in a
nearly-levitating upflow of air obey the Langevin Equation and the
Fluctuation-Dissipation Relation [Ojha {\it et al.} Nature {\bf
427}, 521 (2004) and Phys.\ Rev.\ E {\bf 71}, 016313 (2005)].  To
probe the range of validity of this statistical mechanical
description, we perturb the original experiments in four ways.
First, we break the circular symmetry of the confining potential
by using a stadium-shaped trap, and find that the velocity
distributions remain circularly symmetric.  Second, we fluidize
multiple spheres of different density, and find that all have the
same effective temperature. Third, we fluidize two spheres of
different size, and find that the thermal analogy progressively
fails according to the size ratio.  Fourth, we fluidize individual
grains of aspherical shape, and find that the applicability of
statistical mechanics depends on whether or not the grain chatters
along its length, in the direction of airflow.
\end{abstract}



\maketitle



\section{Introduction}

There is a growing list of driven, far-from-equilibrium systems
where the dynamics of microscopic fluctuations are characterized
by an effective temperature.  One of the earliest examples is the
kinetic energy associated with velocity fluctuations in a sheared
granular material \cite{Bagnold}.  More recent examples in
granular physics include dilute grains driven within a horizontal
plane\cite{Pouligny, Ippolito, Prentis, Rajesh, RajeshPRE05}, as
well as flowing granular liquids \cite{MenonSci97, Mueth, Losert,
Reydellet, PAL00, SongMakse05} and vertically-vibrated granular
gasses \cite{Olafsen, Feitosa, dAnna, Baxter}.  Wider ranging
examples include chaotic fluids \cite{Hohenberg, Egolf}, spin
glasses \cite{CugKur, Herisson}, glasses \cite{Grigera, Berthier},
colloids \cite{Segre, Bellon}, and foams \cite{Ono}, which are all
far away from equilibrium.  In some of these cases \cite{Prentis,
Rajesh, SongMakse05, dAnna, Baxter, Ono}, the behavior is in
perfect analogy with that expected for a system in thermal
equilibrium.  Degrees of freedom are populated according to a
density of states and Boltzmann factor, and correlation-response
relations all hold, with a single effective temperature whose
value is set by the nature of the energy injection mechanism. In
other cases, such a thermal analogy is more limited and does not
hold in detail; for example, the distributions may not be
described by a Boltzmann factor or the effective temperature may
not be uniquely defined.

An outstanding question is how to predict whether or not the
thermal analogy holds. What do the systems in Ref.~\cite{Prentis,
Rajesh, RajeshPRE05, SongMakse05, dAnna, Baxter, Ono} have in
common, and how do they differ from other driven systems? Here we
seek insight by systematically perturbing one case for which the
analogy unarguably holds in all detail, in hopes that it may be
progressively upset. In particular, we focus on a small number of
grains fluidized in a nearly-levitating upflow of air. While
grains thus never leave the plane, they can nevertheless be driven
randomly within the plane by the random shedding of turbulent
wakes at a rate set by the Strouhal number~\cite{Achenbach,
Suryanarayana}. The Reynold's number based on sphere size is of
order $10^4$. Under these conditions, a single sphere confined
within a circular cell rolls stochastically, without slipping,
exactly like a Brownian particle in a two-dimensional harmonic
trap \cite{Rajesh}. Specifically, the dynamics obey a Langevin
equation where the random force autocorrelation is proportional to
the viscous drag memory kernel and the effective temperature
according to the Fluctuation-Dissipation Relation. For a variety
of conditions, the root-mean-squared displacement of the sphere
from the center of the trap and the mean-squared speed of the
sphere, respectively, are given by~\cite{RajeshPRE05}
\begin{eqnarray}
    \sqrt{\langle r^2 \rangle} &=& (0.20\pm0.02)R_{cell},
            \label{Rrms} \\
    \langle v^2\rangle &=&
        0.7\left({\rho_{air}\over{\rho_e}}\right)
        {u^3\over\sqrt{gD}}.
            \label{V2}
\end{eqnarray}
Here $\rho_e=m_e/[(4/3)\pi(D/2)^3]$; $m_e=m+I/(D/2)^2$ is the
effective inertial mass of the sphere; $m$, $I$, and $D$ are
respectively the mass, moment of inertia and diameter of the
sphere; $\rho_{air}$ and $u$ are respectively the density and flow
speed of the air; $g=980~{\rm cm/s^2}$ is gravitational
acceleration; and $R_{cell}$ is the radius of the sample cell.
Physically, Eq.~(\ref{V2}) can be understood by balancing energy
input, via collision between the sphere and a sphere-sized volume
of air, with energy dissipation via viscous drag. Geometrically,
Eq.~(\ref{Rrms}) can be understood by a picture of the repulsion
between the cell wall and the turbulent wake, which expands as it
moves downstream.

The detailed thermal analogy for the behavior of one and two
nearly-levitated gas-fluidized spheres was completely unexpected.
In this paper, we seek insight via systematic perturbation of the
original experiment. To begin, we first describe the experimental
apparatus and analysis procedures used throughout. In the next
four sections, we describe the perturbations and results, with one
perturbation per section.  We shall demonstrate that the thermal
analogy is very robust with respect to some of these
perturbations.  We also shall demonstrate a control parameter by
which the thermal analogy may be progressively upset.

\section{Experimental Details}

Our methods for fluidizing grains and tracking their positions are
similar to those of Refs.~\cite{Rajesh, RajeshPRE05}, but with
some embellishments that we describe in detail here.  As before,
the heart of the apparatus is a rectangular windbox,
$1.5\times1.5\times4~{\rm ft}^3$, standing upright. A circular
sieve with mesh-size 300~$\mu$m sits in a twelve-inch circular
hole on the top. The sieve is horizontal, so that the grains feel
no component of gravity within the plane of motion and so that the
air flow is upward counter to gravity.  Except in the final
section, all grains are spherical and roll without slipping. The
rotational motion is therefore coupled to the translational motion
and can be accounted for by an effective inertial mass as in
Eq.~(\ref{V2}). A digital CCD video camera [Pulnix 6710, 8 bits
deep, 120~Hz frame rate], and a ring of six 100~W incandescent
lights, are located approximately three feet directly above the
sieve, mounted to a scaffolding which in turn is mounted to the
windbox. A blower is connected at the base of the windbox to
provide an upward air flow perpendicular to the sieve.  A hotwire
anemometer measures the flow rate, and verifies its uniformity.
Previously, a perforated metal sheet was fixed in the middle of
the windbox to break up large scale structures in the airflow. To
ensure even more uniformity, we now use two perforated metal
sheets with a one-inch thick foam air-filter sandwiched in
between.

The control of the camera, and all image processing, is
accomplished within LabVIEW. In all runs, images are harvested at
120 frames per second and written to hard disk for post
processing. To minimize the size of the data set, and hence to
optimize the maximum possible run length, we first threshold the
images to binary so that each grain appears as a white blob on a
black background.  The illumination and thresholding level are
adjusted so that each blob corresponds closely to the entire
projected area of the grain. Successive binary images are encoded
as a lossless format AVI movie [Microsoft RLE]. Previously, we
used a custom encoding scheme that is optimal only for a very
small number of spherical grains. The AVI format requires more
disk space, but is also more flexible for large numbers of grains.

\begin{figure}
\includegraphics[width=3.30in]{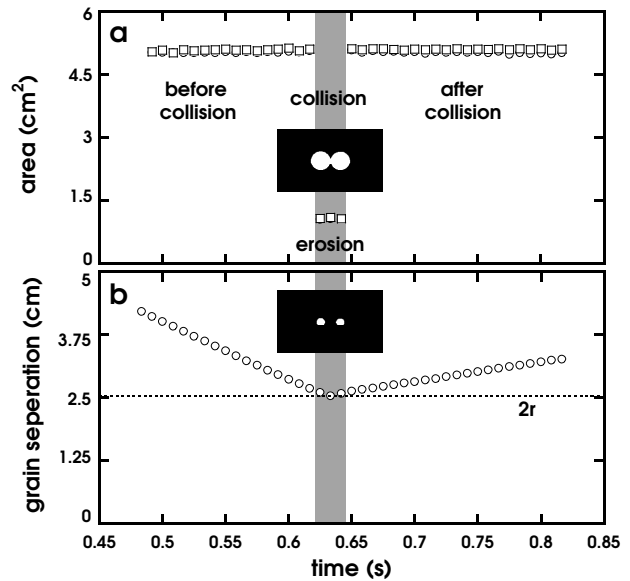}
\caption{[col-ero] (a) The measured blob area, and (b) the
center-to-center separation, for two grains in collision. The
grayed stripe indicates the time-interval the grains are in close
enough proximity so that they are imaged as as a single blob
(inset a), which is eroded producing two smaller separated circles
(inset b).} \label{col-ero}
\end{figure}

For post-processing we also use LabVIEW.  If the grains are far
enough apart as to be distinct white blobs that are completely
surrounded by black, then we use LabVIEW's ``IMAQ Particle
Analysis'' algorithm to locate the center of brightness of each
blob to sub-pixel accuracy. However, when two grains collide and
their blobs touch, then this algorithm identifies only the center
of the combined two-grain blob.  When the total number of
identified blobs falls below the known number of grains, we must
modify our tracking procedure. The widely-used technique of
Ref.~\cite{Crocker} cannot be invoked, because it requires the
grain separation be large compared to grain size.  Instead, we
apply an erosion algorithm, in which a square mask consisting of
ones (white) and zeros (black) is run over the binary image.  The
output at each pixel is one (white) if {\it all} the image pixels
under the white region of the mask are white; otherwise, the
output is zero (black).  For spherical grains, we choose a mask
that is about 2/3 the grain size, that is white throughout the
largest inscribed circle, and that is black outside this region.
This construction preserves the circular shape of the blobs, while
eroding their size.  It also separates blobs that are in contact,
and optimizes their circularity after separation. After applying
such an erosion, we then invoke the same centroid-finding
algorithm as before.  These procedures are demonstrated in
Fig.~\ref{col-ero}, which shows two grains before, during, and
after collision.

There are two more steps.  First, the grain coordinates measured
in each frame must be identified with the correct corresponding
grains in the previous frame.  This is aided by the fast frame
rate of our camera, which is such that the maximum displacement in
one frame is much less than the grain size.  Finally, position vs
time data is fitted to a third order polynomial within a window of
$\pm5$ points in order both to smooth and to differentiate to
second order. Gaussian weighting that is nearly zero at the edges
is used to ensure continuity of derivatives.  The rms deviation of
the raw data from the polynomial fit is 0.001~cm, which we take as
an estimate of position accuracy.  This and the frame rate give an
estimate of speed accuracy as 0.1~cm/s.  Indeed these numbers
correspond to a visual inspection of the level of noise in time
traces.

\section{One sphere in a stadium}

\begin{figure}
\includegraphics[width=3.30in]{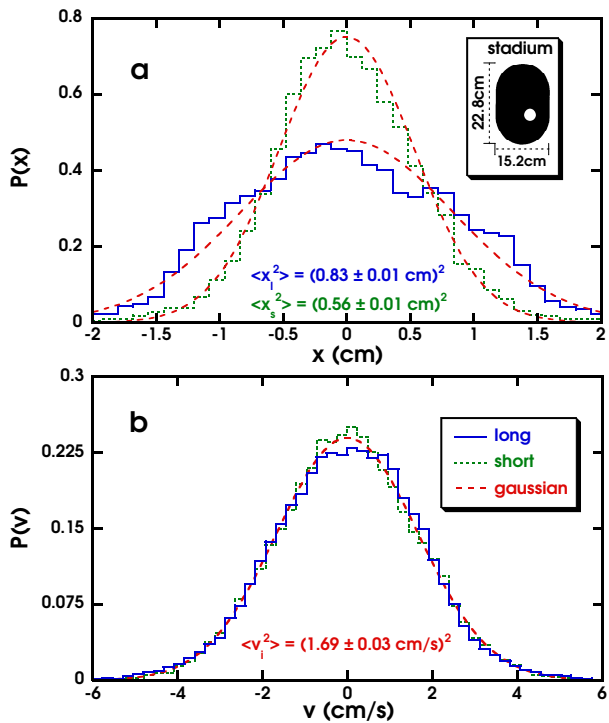}
\caption{[v-p-dis] Distributions for the components of (a)
position and (b) velocity measured for a one-inch nylon sphere
along the long and short axes of a stadium-shaped trap. To within
measurement uncertainty, these distributions are all Gaussian
(dashed curves).} \label{v-p-dis}
\end{figure}

Our first perturbation is motivated by the very form of
Eq.~(\ref{Rrms}), which says that the rms position of the sphere
is set by the radius of the sieve.  So instead of using a circular
sample cell, we now construct a stadium-shaped sample cell by
placing appropriate wooden inserts into contact with the sieve
both above and below. A binary image of our stadium, with a
$D=1$~inch nylon sphere, is shown in the inset of
Fig.~\ref{v-p-dis}. Certainly the elongated boundary will affect
the confining ball-wall potential, with the sphere expected to
move farther along the long axis.  Due to loss of symmetry, the
rms position and speed of the grain could now both be different
along the long and short axes, which would be a direct violation
of the thermal analogy. To investigate, we fluidize the nylon ball
with an upflow of air at speed $u=750$~cm/s, and we track its
position with the methods described above.

Results for the position and speed distributions along the two
axes of the stadium are displayed in Fig.~\ref{v-p-dis}. All four
have the same Gaussian shape, characteristic of a Brownian
particle in a harmonic trap, as seen before.  Though it couldn't
have been expected, the speed distributions remain identical along
the two axes.  However, the sphere now has wider excursions along
the longer axis. The observed rms displacements are $\sqrt{\langle
{x_{l}}^2\rangle} = 0.83$~cm and $\sqrt{\langle {x_{s}}^2\rangle}
= 0.56$~cm for long and short axes, respectively.  The ratio of
these displacements is 1.48, which is very close to the ratio of
long to short dimensions of the stadium, (22.8~cm)/(15.2~cm)=1.50,
in accord with the scaling of Eq.~(\ref{Rrms}).

\begin{figure}
\includegraphics[width=3.30in]{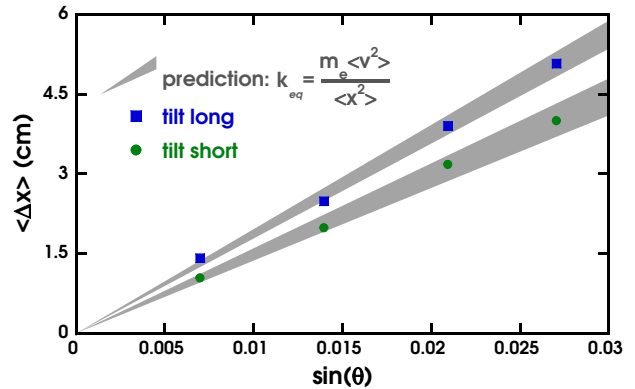}
\caption{[spr-cons] Shift in average position of a fluidized
sphere with respect to the sine of the angle by which the entire
apparatus is tilted.  Data are shown by symbols, and prediction
based on the equipartition assumption and the variance of the
position and speed distributions are shown by shaded regions.  The
experimental conditions are the same as in
Fig.~\protect\ref{v-p-dis}.}\label{spr-cons}
\end{figure}

If the thermal analogy holds for both the position and momentum
degrees of freedom, then the spring constants along the two axes
can be deduced from the equipartition of energy:
\begin{equation}
    T\ = m_e\langle {v_l}^2 \rangle = m_e\langle {v_s}^2 \rangle
    = k_l\langle {x_l}^2 \rangle = k_s\langle {x_s}^2 \rangle,
\label{equip}
\end{equation}
where $T$ is the effective temperature measured in units of
energy.  To test this relation, we compare with an auxiliary
mechanical measurement of the spring constants.  As in
Ref.~\cite{Rajesh}, we tip the entire apparatus by a small angle
$\theta$ away from horizontal and measure the shift $\langle
\Delta x\rangle$ in the average position of the sphere down the
plane. The new average position is where the spring force balances
the force of gravity acting within the plane:
\begin{equation}
    k\langle \Delta x\rangle = mg\sin\theta.
\label{mgsine}
\end{equation}
This is done for orientations of the stadium with the long axis
both parallel and perpendicular to the tilting direction.  The
results for the shift in average position are plotted as symbols
vs the sine of the tilt angle in Fig.~\ref{spr-cons}.  The
expectations based on Eq.~(\ref{equip}) and the position and speed
statistics are also plotted in Fig.~\ref{spr-cons}, now as a
shaded region that reflects measurement uncertainty in the rms
displacements and speeds.  Indeed, the two results agree well.

\begin{figure}
\includegraphics[width=3.30in]{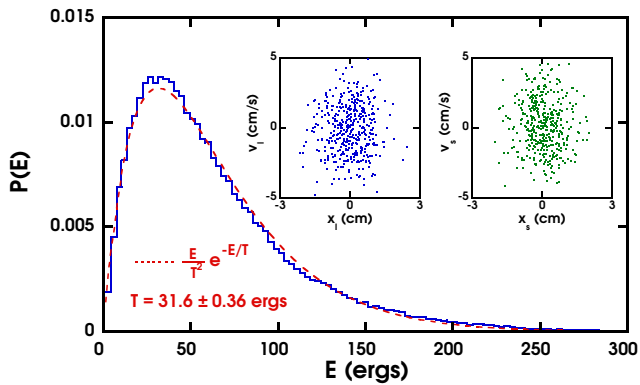}
\caption{[ene-dis] The energy distribution for a one-inch nylon
sphere in a stadium-shaped trap, under the same conditions as in
Figs.~1-2.  The dashed curve shows the expectation based on the
known effective temperature and the product of density of states
times a Boltzmann factor.  The insets show scatter plots of
velocity and position components; successive points are separated
by several seconds, longer than the decay time of the velocity
autocorrelation function.}\label{ene-dis}
\end{figure}

To drive home the validity of the thermal analogy for a
nearly-levitated sphere in a stadium-shaped cell, we now compute
the total mechanical energy $E$ as the sum of kinetic and
potential energies at each instant in time.  We then compute the
distribution of total energy sampled over the entire run.  The
data, shown in the main plot of Fig.~\ref{ene-dis}, agree nicely
with the expectation for a thermal system, $P(E) = (E/T^2)
\exp(-E/T)$, which is given by the density of states times the
Boltzmann factor with no adjustable parameters.  The insets show
no correlation in phase-space scatter plots of speed vs position.

While we might have hoped to tune the validity of the thermal
analogy by the value of the aspect ratio of the sample boundary,
apparently it is robust with respect to this perturbation and we
must look elsewhere.

\section{Five spheres of different density}

Our second perturbation is motivated by the form of
Eq.~(\ref{V2}), which specifies the mean-squared speed of
individually fluidized spheres as a function of the air and sphere
properties. Note that Eq.~(\ref{V2}) gives the scaling of the mean
kinetic energy with sphere density as $K\sim 1/\rho_e$, due to the
way energy is injected by turbulent wakes.  If this relation also
holds when multiple spheres of different density are in the sample
cell, then the spheres would have different temperatures. To test
this possibility we now simultaneously fluidize five solid spheres
of the same diameter, $D=2.54$~cm, but of different density. The
materials and effective densities of the spheres are as follows:
wood 0.95~g/cc; polypropylene 1.29~g/cc; nylon 1.57~g.cc; teflon
3.02~g/cc; Al$_2$O$_3$ ceramic 5.33~g/cc.  We note that the wooden
ball is slightly aspherical, and its diameter is about 0.5\%
smaller than the others.  The air speed is $u=600$~cm/s, the trap
is circular, and the sieve is perpendicular to gravity.

\begin{figure}
\includegraphics[width=3.30in]{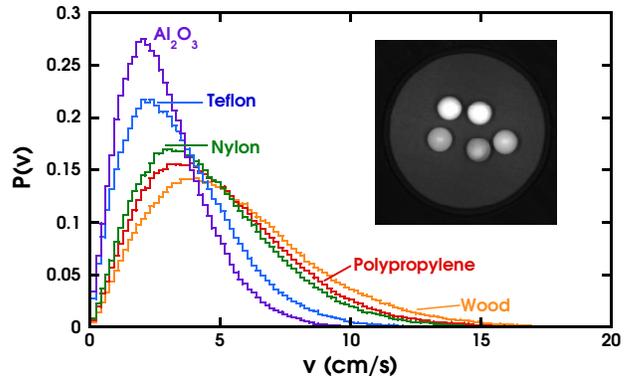}
\caption{[diffdens] Speed distributions for grains of equal
diameter but different density. In order of decreasing density,
the sphere materials are ceramic (Al$_2$O$_3$), Teflon, Nylon,
polypropylene, and wood.  The inset shows a photograph of the
system.}\label{diffdens}
\end{figure}

\begin{figure}
\includegraphics[width=3.30in]{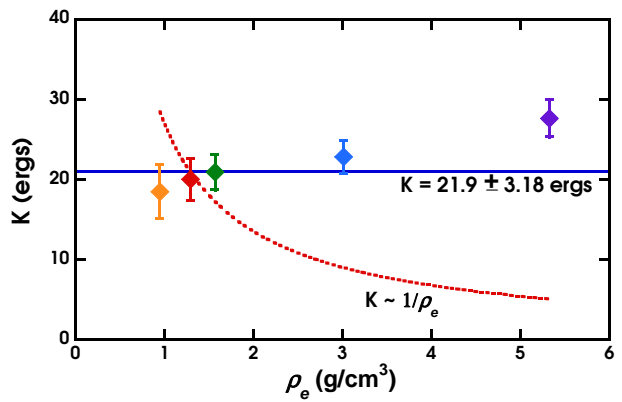}
\caption{[ener] Average kinetic energy vs density, based on the
speed distributions shown in Fig.~\protect\ref{diffdens} for five
spheres of equal diameter. To within measurement uncertainty, the
results are nearly constant, as shown by the solid horizontal
line.  The expectation based on single-grain behavior,
Eq.~(\protect\ref{V2}), is shown by a dashed curve.} \label{ener}
\end{figure}

The normalized speed distributions are displayed in
Fig.~\ref{diffdens} for the five spheres.  All have the same
Gaussian form as for a thermal particle in two dimensions,
$P(v)=(2v/\langle v^2\rangle)\exp(-v^2/\langle v^2\rangle)$.  But
evidently the lighter spheres move faster, on average, than the
denser spheres.  The mean kinetic energy for each sphere is
plotted in Fig.~\ref{ener} vs effective density.  There is a
slight upward trend, with a difference of about thirty percent
from lightest to densest spheres.  This rise is slightly larger
that the measurement uncertainty.  More crucially, it is also
slight in comparison with the factor of five decrease predicted by
Eq.~(\ref{V2}) for one ball alone, $K\sim1/\rho_e$; Eq.~(\ref{V2})
can be completely ruled out for multi-ball systems.  Apparently
the spheres exchange energy, mainly through interaction of their
wakes as well as through occasional direct collisions, and thereby
come to almost the same temperature. The thermal analogy is fairly
robust with respect to perturbation of sphere density.

\section{Two spheres of different size}

Our next perturbation is also motivated by the form of
Eq.~(\ref{V2}), which implies that the mean kinetic energy of an
individually fluidized sphere scales with sphere diameter as
$K\sim D^{5/2}$.  If this holds when multiple spheres of different
diameter are simultaneously fluidized, then the spheres would have
different temperatures.  To test this possibility, we fluidize
pairs of nylon spheres of different diameter.  The airspeed is
$u=770$~cm/s and the trap is circular.  By varying the choice of
spheres, we have examined the behavior for seven diameter ratios
ranging from about 0.5 to 5.

\begin{figure}
\includegraphics[width=3.30in]{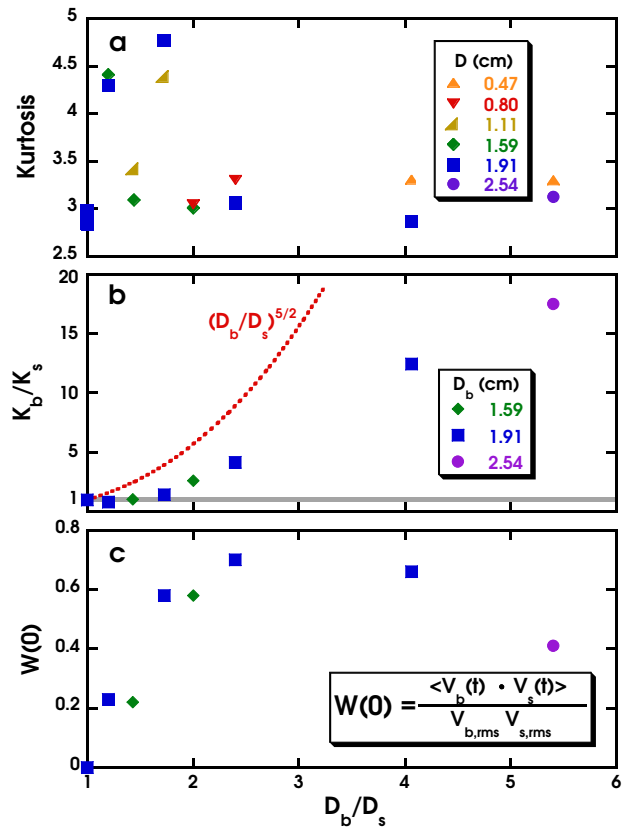}
\caption{[TRK] (a) Kurtosis of the velocity distributions, (b)
ratio of average kinetic energies, and (c) equal-time velocity
cross-correlation for pairs of nylon spheres vs the ratio of their
diameters.  For the thermal analogy to hold, the kurtosis should
equal 3, the kinetic energy ratio should equal 1, and the
equal-time velocity cross-correlation should vanish.  The ratio of
average kinetic energies predicted by single-sphere behavior,
Eq.~(\ref{V2}), is shown by a dashed curve.}\label{TRK}
\end{figure}

The speed distributions are always nearly Gaussian. This is
quantified in Fig.~\ref{TRK}a, which shows the kurtosis $\langle
v^4 \rangle / \langle v \rangle^4$ of the speed distribution for
each sphere as a function of diameter ratio.  The values are close
to 3, the Gaussian expectation, except for two cases.  This in and
of itself is a violation of the thermal analogy.  However, it is
not so drastic that the mean kinetic energy, and hence the
effective temperature, become ill-defined.

The ratio of mean kinetic energies of the two balls, which equals
the ratio of their effective temperatures, is plotted vs diameter
ratio in Fig.~\ref{TRK}b.  The kinetic energies are nearly equal
for diameter ratios of less than two.  But for increasing diameter
ratio, the larger sphere becomes progressively hotter than the
smaller sphere.  Evidently, the diameter ratio is a control
parameter that can be varied to systematically break the thermal
analogy.  This breakdown appears to be quite gentle, though. The
temperature ratio is not as great as expected by Eq.~(\ref{V2}),
which again we find to be incorrect for multi-ball systems. Also,
the leading behavior is not linear, but rather quadratic in the
diameter ratio.  This may be amenable to theoretical modelling.

Before closing this section, we now consider the physical origin
of the breakdown of the thermal analogy vs diameter ratio.  The
reason, actually, is immediately obvious when viewing the system
directly. The two spheres usually repel one another through
interaction of their wakes, as discussed in
Ref.~\cite{RajeshPRE05}. However, if they approach close enough,
then they come into lasting contact and the upflow of air exerts a
net total force on the pair causing them to accelerate straight
across the cell until reaching the boundary. The direction of
motion is such that the large sphere appears to chase the small
sphere out of its territory.  We speculate that the loss of
symmetry of the two-ball pair causes the vortices to be shed
preferentially along the line of centers, resulting in a net
force.

The prevalence of this ``chasing''phenomenon may be quantified by
the equal-time velocity cross-correlation, $\langle {\bf v_b}(t)
\cdot {\bf v_s}(t) \rangle$, where the subscripts denote ``big''
and ``small'' spheres as before.  For the thermal analogy to hold,
this quantity must vanish because all kinetic degrees of freedom
must be independently populated.  During a chase, however, the two
velocities are equal and hence perfectly correlated.  Data for the
equal-time velocity cross-correlation, made dimensionless by the
rms speeds of the two balls, are plotted vs diameter ratio in
Fig.~\ref{TRK}c.  By contrast with the effective temperature
ratio, this rises abruptly from zero for diameter ratios greater
than one.  Also by contrast, it reaches a peak for a diameter
ratio of about 3 and then decreases.  If the size disparity is too
small, then the loss of symmetry is not enough to cause much
chasing.  If the size disparity is too great, then the large grain
slowly rolls without regard for the small ball, which quickly
flits about and is repelled as though from a stationary object. We
believe the thermal analogy is recovered in this limit, but with
the two balls being differentially ``heated'' by the upflow of
air.

\section{One aspherical grain}

\begin{figure}
\includegraphics[width=3.30in]{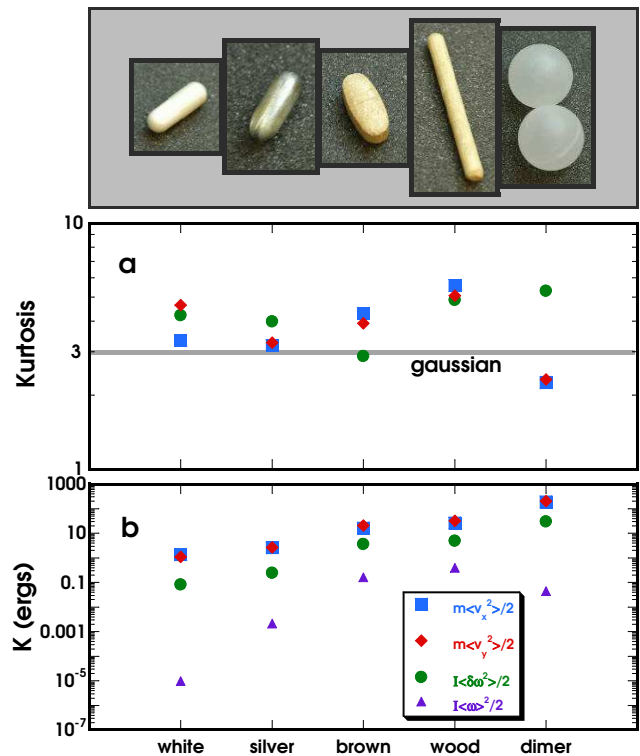}
\caption{[AllShape] (a) Photographs of five aspherical grains; (b)
the kurtosis of the speed distributions, and (c) the average
kinetic energies, for the two translational and the one rotational
degrees of freedom when these grains are individually fluidized.
The airspeeds are 910~cm/s for the three leftmost grains, and
750~cm/s for the two rightmost grains.}\label{AllShape}
\end{figure}

In the above sections, and also in Refs.~\cite{Rajesh,
RajeshPRE05}, the shape of the grain is spherical.  This is
special because it permits the grain to roll freely in all
directions without sliding.  It is also special because it permits
vortices to be shed equally in all directions.  To explore for
qualitatively new phenomena, and to seek another means of
violating the thermal analogy, we now perturb the grain shape. The
objects we fluidize are listed in Table~\ref{aspherical} and
pictured in Fig.~\ref{AllShape}: various pharmaceutical pills, a
cylindrical wooden rod, and a dimer consisting of two connected
hollow plastic spheres. When individually fluidized by an upflow
of air, the pictured grains all translate and rotate seemingly at
random. A few of the grains are axisymmetric, like the dimer;
however, they exhibit virtually no rotation about the axis of
continuous symmetry. While the spheres in previous sections roll
without slipping, here the aspherical grains must slide in order
to translate or rotate.

We may characterize the motion of these grains in terms of the
time dependence of their center-of-mass position and their angular
orientation.  The former is deduced as per the spherical grains
from the center of brightness.  The latter is deduced from second
moments of the spatial brightness distribution.  Then we
differentiate to measure both translational and rotational speed
distributions.  None of the grains is chiral by design, since that
would lead to steady whirling in one direction.  Nonetheless, some
whirling can occur due to imperfections in shape.  Therefore we
measure both the average angular speed $\langle \omega\rangle$ as
well as fluctuations $\delta\omega$ about this average.

\begin{table}[ht]
\begin{ruledtabular}
\begin{tabular}{lcccccc}
Name & $\rho$ (g/cc) & $L$ (cm) & $W$ (cm) & $H$ (cm) & $m$ (g) &
$I$ (g-cm$^2$) \\ \colrule
white  & 0.685 & 2.12 & 0.848 & 0.848 & 0.82 & 0.344 \\
silver & 0.951 & 1.52 & 0.586 & 0.586 & 0.39 & 0.083 \\
brown  & 0.937 & 1.94 & 0.966 & 0.888 & 1.56 & 0.611 \\
wood   & 0.671 & 4.75 & 0.540 & 0.540 & 0.73 & 1.386 \\
dimer  & 0.256 & 5.08 & 2.540 & 2.540 & 4.40 & 11.34 \\
\end{tabular}
\end{ruledtabular}
\caption{The density, length, width, height, mass, and moment of
inertia of five aspherical grains.  For computation of density and
moment of inertia, the white, silver, and wood grains are
approximated as cylinders, while the brown grain is approximated
as a block. The dimer is composed of two polypropylene shells of
thickness 0.14~cm.} \label{aspherical}
\end{table}

A summary of the results for all grains, individually fluidized,
is shown in Fig.~\ref{AllShape}.  The kurtosis of the
translational and rotational speed distributions is shown in the
top plot.  The results appear statistically greater than 3, the
gaussian result, except for the translational velocity components
of the dimer. The average kinetic energies are shown in the bottom
plot.  They too exhibit a violation of the thermal analogy since
the translational kinetic energy is greater than the rotational
kinetic energy.  At this airspeed, the energy associated with
whirling, $I\langle\omega\rangle^2 /2$, is at least ten times
smaller than the energy of angular speed fluctuations, $I\langle
\delta\omega^2 \rangle/2$.  Therefore, the whirling caused by
slight shape imperfections is not responsible for the breakdown of
the thermal analogy.

\begin{figure}
\includegraphics[width=3.30in]{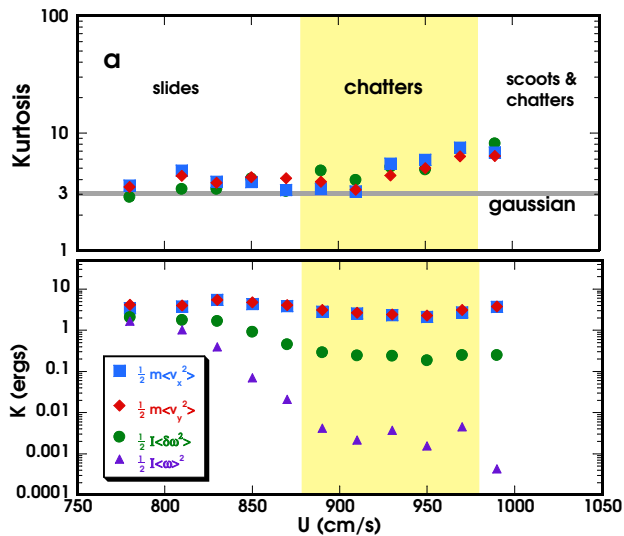}
\caption{[Silver] (a) The kurtosis of the speed distributions, and
(b) the average kinetic energies, for the two translational and
the one rotational degrees of freedom for the silver grain in
Fig.~\protect\ref{AllShape}, as a function of the fluidizing
airspeed.}\label{Silver}
\end{figure}

To systematically explore the range of behavior for aspherical
grains, we now vary the airflow for just one shape.  We choose the
silver pill, for which the thermal analogy works best in
Fig.~\ref{AllShape}. Results for the average energy in each of the
three kinetic degrees of freedom, $m \langle v_x^2\rangle/2$,
$m\langle v_y^2\rangle/2$, and $I\langle\delta\omega^2\rangle/2$,
as well as the whirling energy $I\langle\omega\rangle^2/2$, are
shown in Fig.~\ref{Silver} along with the kurtosis of the
distributions. Counter to intuition, and also counter to
Eq.~(\ref{V2}), the translational kinetic energy is nearly
constant while the rotational kinetic energy actually decreases
with increasing airspeed.  As airspeed decreases, the kurtosis
values decrease towards three and both the whirling and rotational
fluctuation energies approach the translational kinetic energies.
Except for the whirling, the motion is more nearly thermal at
lower airspeeds.

The sequence of behavior in Fig.~\ref{Silver} correlates with the
motion of the grain {\it perpendicular} to the sieve, which cannot
be captured by our usual video methods.  At low airspeeds, the
grain is in physical contact with the sieve; translational motion
thus requires sliding. At intermediate airspeeds, the center of
mass is raised somewhat and the grain chatters back and forth
along its length. This chattering becomes more prevalent as the
airspeed increases. At the highest airspeeds, the chattering
motion continues but with the important difference that
occasionally the grain scoots rapidly across the cell. This is
somewhat reminiscent of the intermittent chasing observed for two
spheres of different size, and it too ruins the thermal analogy.
Perpendicular motion is important for the other aspherical grains,
as well. At the given airspeeds in Fig.~\ref{AllShape}, the white
and silver grains both chatter steadily. The brown grain, wooden
rod, and dimer all slide without chattering, like the silver grain
at low airspeeds. To fully characterize and understand the
behavior of aspherical grains, it would be necessary to measure
their out-of-plane motion.

\section{Conclusion}

In summary we have explored four systematic perturbations to an
experiment on nearly-levitated spheres that was previously
\cite{Rajesh, RajeshPRE05} discovered to behave in perfect analogy
to a thermal system.  Here we find that the statistical mechanical
description is robust with respect to variation of both the shape
of the sample cell and with respect to the densities of the
spheres.  This adds to the growing list of driven
out-of-equilibrium systems for which an effective temperature may
be defined and used in the usual statistical mechanical sense.
However, we also find that the spheres must have the same diameter
or else the thermal analogy progressively breaks down as the size
disparity increases. Furthermore, the analogy is well-controlled
only for spherical grains.  It can work for pill-shaped objects,
but depends on out-of-plane motion that has not yet been well
characterized.  We hope that the smooth, gradual breakdown as a
function of diameter ratio will stimulate theoretical work.  This
could lead to a better general understanding of when the concepts
and tools of statistical mechanics can be invoked for driven
far-from-equilibrium systems.

\begin{acknowledgments}
We thank Rajesh Ojha and Paul Dixon for their help. This work was
supported by NSF through Grant No.~DMR-0514705.
\end{acknowledgments}

\bibliography{AirballRefs}

\end{document}